\documentclass[pre,twocolumn,aps,10pt]{revtex4-2}
\usepackage{amsmath}
\usepackage{amssymb}
\usepackage{graphicx}
\usepackage{braket}
\usepackage{stmaryrd}
\usepackage{hyperref}
\usepackage{color}
\usepackage{dsfont}
\usepackage{bm}

\hypersetup{
    colorlinks,
    citecolor=blue,
    linkcolor=blue,
    urlcolor=blue
}

\begin{document}

\title{Unified entropy production in finite quantum systems
}
\author{Tomohiro Nishiyama}
\email{htam0ybboh@gmail.com}
\affiliation{Independent Researcher, Tokyo 206-0003, Japan}

\author{Yoshihiko Hasegawa}
\email{hasegawa@biom.t.u-tokyo.ac.jp}
\affiliation{Department of Information and Communication Engineering, Graduate
School of Information Science and Technology, The University of Tokyo,
Tokyo 113-8656, Japan}

\date{\today}
\begin{abstract}

In finite-dimensional quantum systems, temperature cannot be uniquely defined.
This, in turn, implies that there are several ways to define entropy production in finite-dimensional quantum systems, because the classical entropy production depends on temperature.
We propose a unified definition of entropy production based on the difference in quantum relative entropy with respect to reference states characterized by effective temperatures.
We demonstrate that the proposed definition naturally decomposes into a Clausius-type entropy production and an additional contribution arising from the time dependence of the effective temperature.
Furthermore, we show that requiring the entropy production rate to take the conventional form as the sum of the entropy change and the heat flow constrains the effective temperature to be either constant or equal to a specific energy-matching effective temperature.
For general initial states, entropy production can become negative, in which case we derive lower bounds on entropy production and establish sufficient conditions for its non-negativity using the trace distance.
\end{abstract}
\maketitle

\section{Introduction\label{sec:Introduction}}

Temperature is a fundamental concept in thermodynamics;
however, it is usually defined for equilibrium systems, and defining temperature
in nonequilibrium settings, particularly for finite systems, remains highly nontrivial~\cite{CasasVazquez:2003:NoneqTemperature}.
In the quantum regime, several notions of effective temperature have been proposed depending on the operational or theoretical context.
For instance, one may define a time-dependent temperature through an entropy--energy relation (e.g., via $dS/dE$)~\cite{alipour2021temperature},
or define temperature operationally depending on the chosen task or measurement protocol~\cite{lipka2023operational}.
Related approaches have also been discussed in the context of finite environments and trajectory-based formulations~\cite{moreira2023stochastic}.
Various operational and theoretical definitions have been proposed~\cite{alipour2021temperature, lipka2023operational, moreira2023stochastic, strasberg2021first, strasberg2021clausius}.
The lack of a unique temperature in nonequilibrium systems is problematic because many thermodynamic quantities depend on temperature.
One notable example is the entropy production, which quantifies the irreversibility of the dynamics. 
The entropy production plays a fundamental role in quantifying the thermodynamic cost of thermal machines and in thermodynamic uncertainty relations~\cite{Barato:2015:UncRel,Gingrich:2016:TUP,Garrahan:2017:TUR,Hasegawa:2019:FTUR,Erker:2017:QClockTUR,Carollo:2019:QuantumLDP,Guarnieri:2019:QTURPRR,Hasegawa:2020:QTURPRL}.
In quantum thermodynamics, several definitions of entropy production have been proposed and reviewed~\cite{Landi:2021:EPReview}.
For instance, entropy production can be expressed in terms of a relative entropy between the system-environment state and an appropriate reference state~\cite{Esposito:2010:EntProd},
or defined at the level of stochastic trajectories in quantum jump/trajectory formalisms~\cite{Manzano:2018:EPPRX}.
For finite environments, a Clausius-type entropy production based on an effective inverse temperature has been proposed~\cite{strasberg2021clausius}.
Usually, the quantum entropy production is defined as the sum of the change in the system von Neumann entropy and 
the energy change in the environment divided by the temperature.
As mentioned above, since the temperature is not uniquely defined in finite-dimensional environments, there is no consensus on the definition of entropy production in finite-dimensional quantum systems.

In this paper, we propose a unified definition for entropy production with a general effective temperature [cf. Eq.~\eqref{eq:def_EP_geometric}]. In defining entropy production, we require the following three conditions: 1) when the initial state is in product form and the initial state of the environment is taken as the Gibbs state, the entropy production is non-negative (the second law of thermodynamics), 2) when the initial and the final state including effective temperatures are identical, the entropy production
becomes zero, 3) the definition reduces to the definition for the large environment when the effective temperature is a constant. The simplest definition that satisfies these three conditions is given via the difference of the quantum relative entropy. We show that our definition reduces to the previously proposed Clausius-type entropy production for the effective temperature $\beta^{\ast}_t$,  which is defined such that the instantaneous mean energy of the environment agrees with that of a Gibbs state [cf. Eq.~\eqref{eq:inverse_temp_condition}].   Moreover, we show that for the entropy production rate to correspond to the sum of the change in entropy and the change in heat, the effective temperature is restricted to either a constant or $\beta^{\ast}_t$.
Since the entropy production can be negative for general initial states, even for a constant temperature, as also demonstrated experimentally~\cite{aimet2025experimentally}, we derive lower bounds for the entropy production with effective temperature.
In this context, generalized second-law-type bounds (related to the Landauer principle) have been discussed for negative entropy production~\cite{mondal2023modified}.
Based on our bounds, we further provide sufficient conditions for non-negative entropy production.

\section{Methods}
Consider an open quantum system consisting of the system $S$ and the environment $E$ with finite dimension $d_S<\infty$ and $d_E<\infty$.
Let $H_S(t)$ and $H_E$ be the Hamiltonian operators representing the system $S$ and the environment $E$, respectively. Let $H_{SE}(t)$ be the interaction Hamiltonian operator between the system and the environment. We assume that $H_E$ does not vary over time and $H_E$ has at least two distinct eigenvalues.
The total Hamiltonian is defined as 
\begin{align}
    H(t)&=H_{S}(t)+H_{E} +H_{SE}(t).
    \label{eq:def_Hamiltonian}
\end{align}
Let $\rho_{SE}(t)$ be a density operator, and $\rho_{SE}(t)$ evolves under a joint unitary operator $U_{t}=\mathbb{T} e^{-i\int_{0}^{t} H(s)ds}$:
\begin{align}
    \rho_{SE}(t)=U_{t}\rho_{SE}(0) U_{t}^\dagger.
    \label{eq:evolution_density}
\end{align}
Here $\mathbb{T}$ denotes the time-ordering operator. Throughout this paper, we adopt the convention of setting $\hbar =1$ and $k_B=1$. Let $\mathrm{tr}_{X}$ be a partial trace with respect to $X:=\{S,E\}$, and let
$\rho_{S}:=\mathrm{tr}_{E}[\rho_{SE}]$ and $\rho_E:=\mathrm{tr}_{S}[\rho_{SE}]$ be the density operators of the system and the environment, respectively. 
We define the von Neumann entropy for the density operator $\rho$ as 
\begin{align}
    S(\rho):=-\mathrm{tr}[\rho \ln \rho].
\end{align}
In the following, we refer to the von Neumann entropy as entropy. For $\rho_{X}$, the entropy is defined as $S(\rho_X):=-\mathrm{tr}_{X}[\rho_{X}\ln \rho_{X}]$.
Let $ I(\rho_{SE})$ be the mutual information:
\begin{align}
    I(\rho_{SE}):=S(\rho_S) + S(\rho_E)- S(\rho_{SE})\geq 0,
    \label{eq:def_mutual_info}
\end{align}
and let $D(\rho\|\sigma)$ be a quantum relative entropy:
\begin{align}
    D(\rho\|\sigma):=\mathrm{tr}[\rho(\ln \rho-\ln\sigma)].
    \label{eq:def_quantum_RE}
\end{align}
Let $\gamma_E(\beta)$ be a Gibbs state at an inverse temperature $\beta$:
\begin{align}
    \gamma_E(\beta):=\frac{e^{-\beta H_E}}{\mathcal{Z}(\beta)},
\end{align}
where $\mathcal{Z}(\beta):=\mathrm{tr}_E[e^{-\beta H_E}]$ denotes a partition function.

\subsection{Clausius-type entropy production  \label{sec:eff_temp}}

Let $\beta^\mathrm{eff}_t$ be an arbitrary time-varying effective inverse temperature.
For the sake of notation, $\beta^{\mathrm{eff}}_t$ is referred to as $\beta_t$, and $\rho_{SE}(0)$, $\rho_{SE}(\tau)$ are referred to as $\rho_{SE}$ and $\sigma_{SE}$, respectively. Let $\Delta S_{X}:=S(\sigma_X)-S(\rho_X)$ for $X=\{S,E\}$, and let $\Delta I:=I(\sigma_{SE})-I(\rho_{SE})$. We use the notation $F[\beta_t]$ as the functional of the effective inverse temperature $\beta_t$.

Consider the Clausius-type entropy production from $t=0$ to $\tau$:
\begin{align}
    \Delta\Sigma_{\mathrm{CL}}[\beta_{t}]&:=\Delta S_S+\int_0^\tau  \beta_{t} \frac{d}{dt}\mathrm{tr}_E[\rho_E(t)H_E]dt\nonumber\\
    &=\Delta S_S-\int_0^\tau \beta_{t} \frac{dQ(t)}{dt}dt,
    \label{eq:def_EP_time_variant_beta}
\end{align}
where $dQ(t)/dt=-\partial_t\mathrm{tr}_E[\rho_E(t)H_E]$ is the heat flux into the system. Let $\beta^C$ be a constant effective temperature. 
When $\beta_t=\beta^C$, $\Delta\Sigma_{\mathrm{CL}}[\beta_t]$ reduces to the well-known definition: 
\begin{align}
    \Delta\Sigma_{\mathrm{CL}}[\beta^C]&=\Delta S_S+\beta^C(\mathrm{tr}_E[\sigma_E H_E] - \mathrm{tr}_E[\rho_EH_E]).
    \label{eq:EP_const_beta}
\end{align}
Reference.~\cite{Esposito:2010:EntProd, riechers2021initial, mondal2023modified, ptaszynski2023ensemble} showed that, for a constant inverse temperature $\beta^C$, the Clausius-type entropy production $\Delta\Sigma_{\mathrm{CL}}[\beta^C]$ can be written as the change in quantum relative entropy with respect to the reference state $\rho_{S}(t)\otimes \gamma_E(\beta^C)$:
\begin{align}
    \Delta\Sigma_{\mathrm{CL}}[\beta^C]&=D(\sigma_{SE}\|\sigma_{S}\otimes \gamma_E(\beta^C))-D(\rho_{SE}\|\rho_{S}\otimes \gamma_E(\beta^C))\nonumber\\
    &=\Delta I + D(\sigma_E\|\gamma_E(\beta^C))-D(\rho_E\|\gamma_E(\beta^C)).
    \label{eq:EP_D_const}
\end{align}
To our knowledge, this representation was first explicitly stated in Ref.~\cite{riechers2021initial}, where a time-independent reference state (not necessarily Gibbs) was employed to define entropy production.
The second equality follows from the identity~\cite{Esposito:2010:EntProd}:
\begin{align}
    D(\rho_{SE}(t)\|\rho_{S}(t)\otimes \gamma_E(\beta))=I(\rho_{SE}(t))+D(\rho_E(t)\|\gamma_E(\beta)).
    \label{eq:D_decomp_mutual_DE}
\end{align}
 
Unlike the case of constant temperature, it is not clear whether the entropy production with time-dependent effective temperature should be defined based on the quantum relative entropy or the Clausius-type entropy production. Reference.~\cite{strasberg2021first, strasberg2021clausius} suggested a Clausius-type entropy production for a finite environment:
\begin{align}
     \Delta \Sigma^{\ast}:=\Delta\Sigma_{\mathrm{CL}}[\beta^{\ast}_{t}].
     \label{eq:EP_timevariant_beta}
\end{align}
Here the effective inverse temperature $\beta^{\ast}_{t}$ is defined such that the energy expectation value of $\rho_E(t)$ is equal to the Gibbs state: 
\begin{align}
    \mathrm{tr}_E[\rho_E(t) H_E]=\mathrm{tr}_E[\gamma_E(\beta^{\ast}_{t}) H_E].
    \label{eq:inverse_temp_condition}
\end{align}
This equation always has a unique solution by extending the definition $\beta^{\ast}_{t}$ to negative values (see Appendix~\ref{sec:uniqueness}). Therefore, the effective temperature is well-defined and matches the standard definition when $\rho_E(t)$ is the Gibbs state.

\section{Results}

Before defining entropy production with general effective temperatures, we discuss the conditions that the entropy production must satisfy.
Since definitions of entropy production [Eqs.~\eqref{eq:EP_const_beta} and~\eqref{eq:EP_timevariant_beta}] are known to be non-negative when $\rho_{SE}=\rho_{S}\otimes \gamma_E(\beta^C)$~\cite{Esposito:2010:EntProd} or $\rho_{SE}=\rho_{S}\otimes \gamma_E(\beta^{\ast}_0)$~\cite{strasberg2021first}, these definitions satisfy the second law of thermodynamics for the specific initial state, which is in product form and the initial state of environment is the Gibbs state.
Recall that the total system is closed, if the density operator $\rho_{SE}(t)$ periodically returns to the initial state and the effective temperature is also defined periodically, the entropy production should remain unchanged. 
For time-independent quantum systems with discrete energy eigenstates, the quantum recurrence theorem~\cite{bocchieri1957quantum} guarantees the existence of the aforementioned periodicity by purification of the density operator.

\subsection{Definition of entropy production}
Based on these considerations, we require the following three conditions for the entropy production with the general effective temperature:
\begin{enumerate}
\item  When the initial state is in product form and the initial state of the environment is taken as the Gibbs state (i.e., $\rho_{SE}=\rho_{S}\otimes \gamma_E(\beta_0)$), the entropy production is non-negative (the second law of thermodynamics).
\item When the initial and the final state including effective temperatures are identical (i.e., $\sigma_{SE}=\rho_{SE}$ and $\beta_{\tau}=\beta_0$), the entropy production becomes zero. 
\item The definition reduces to Eq.~\eqref{eq:EP_D_const} when $\beta_t=\beta^C$.
\end{enumerate}
The last condition is introduced to match the definition of entropy production for the large environment.
We propose a definition of the entropy production with effective inverse temperature from time $t=0$ to $\tau$:
\begin{align}
    \Delta \Sigma(\beta_{0},\beta_{\tau})&:=D(\sigma_{SE}\|\sigma_{S}\otimes \gamma_E(\beta_{\tau}))-D(\rho_{SE}\|\rho_{S}\otimes \gamma_E(\beta_0))\nonumber\\
    &=\Delta I + D(\sigma_E\|\gamma_E(\beta_{\tau}))-D(\rho_E\|\gamma_E(\beta_{0})).
    \label{eq:def_EP_geometric}
\end{align}
This is the first main result of this paper.
Equation~\eqref{eq:def_EP_geometric} is the simplest definition that satisfies conditions 1 through 3.
By interpreting $\rho_{S}(t)\otimes \gamma_E(\beta_{t})$ as a reference state approximating the density operator $\rho_{SE}(t)$, the right-hand side represents how the difference between the reference state and the actual state at time $t$ has changed from $t=0$ to $\tau$. Although alternative definition candidates $\Delta\Sigma_{\mathrm{CL}}[\beta_{t}]$ and $\Delta S_S+\beta_{\tau}\mathrm{tr}_E[\sigma_E H_E] - \beta_{0}\mathrm{tr}_E[\rho_EH_E]$ satisfy condition 3, it is not clear whether they satisfy condition 1 or 2.

\subsection{Relation with Clausius-type entropy production}
We show the relation between Eq.~\eqref{eq:def_EP_geometric} and the Clausius-type entropy production $\Delta \Sigma_{\mathrm{CL}}[\beta_{t}]$, and we show that our definition includes the definition Eq.~\eqref{eq:EP_timevariant_beta}. 
The relation between $\Delta \Sigma(\beta_{0},\beta_{\tau})$ and $\Delta\Sigma_{\mathrm{CL}}[\beta]$ is given by
\begin{align}
    \Delta \Sigma(\beta_{0},\beta_{\tau})=\Delta \Sigma_{\mathrm{CL}}[\beta_{t}]+\Delta D[\beta_t],
    \label{eq:relation_clausius}
\end{align}
where 
\begin{align}
    \Delta D[\beta_t]&:=\int_0^\tau \dot{\beta_t} \left.\frac{\partial}{\partial\beta} D(\rho_{SE}(t)\|\rho_{S}(t)\otimes \gamma_E(\beta))\right|_{\beta=\beta_{t}}dt\nonumber\\
    &=\int_0^\tau \dot{\beta_t} (\mathrm{tr}_E[\gamma_E(\beta^{\ast}_t)H_E]-\mathrm{tr}_E[\gamma_E(\beta_t)H_E])dt.
    \label{eq:Delta_D_beta}
\end{align}
Equation~\eqref{eq:relation_clausius} is the second main result of this paper.
The proof is shown in Appendix~\ref{sec:relation_clausius}.
The term $\Delta D[\beta_t]$ represents the change in  $D(\rho_{SE}(t)\|\rho_{S}(t)\otimes \gamma_E(\beta_t))$  caused by the time dependence of the effective temperature.
Since the left-hand side of Eq.~\eqref{eq:relation_clausius} represents the total change of  $D(\rho_{SE}(t)\|\rho_{S}(t)\otimes \gamma_E(\beta_t))$  from $t=0$ to $\tau$, $\Delta \Sigma_{\mathrm{CL}}[\beta_{t}]$ can be interpreted as the change arising solely from the time evolution of the density operator $\rho_{SE}(t)$. 
As a corollary of Eq.~\eqref{eq:relation_clausius}, it follows that the definition Eq.~\eqref{eq:def_EP_geometric} reduces to Eq.~\eqref{eq:EP_timevariant_beta} when $\beta_t=\beta^{\ast}_t$: 
\begin{align}
    \Delta \Sigma(\beta^{\ast}_{0},\beta^{\ast}_{\tau}) &=\Delta \Sigma^{\ast}.
    \label{eq:tilde_Sigma_equality}
\end{align}

We show the relation between $\Delta \Sigma(\beta_{0},\beta_{\tau})$ and $\Delta \Sigma^{\ast}$. 
From the result in~\cite{reeb2014improved}, the Pythagorean identity holds between $\rho_E(t)$, $\gamma_E(\beta^{\ast}_{t})$, and $\gamma_E(\beta)$ (see Appendix~\ref{sec:Pythagorean}):
\begin{align}
    D(\rho_E(t) \| \gamma_E(\beta))=D(\rho_E(t) \| \gamma_E(\beta^{\ast}_{t}))+D(\gamma_E(\beta^{\ast}_{t})\|\gamma_E(\beta)).
    \label{eq:Pythagorean_eq}
\end{align}
Equations~\eqref{eq:D_decomp_mutual_DE} and~\eqref{eq:Pythagorean_eq} yield the Pythagorean identity again as follows.
\begin{align}
    &D(\rho_{SE}(t)\|\rho_{S}(t)\otimes \gamma_E(\beta_t))\nonumber\\
    &=I(\rho_{SE}(t))+D(\rho_E(t) \| \gamma_E(\beta^{\ast}_{t}))+D(\gamma_E(\beta^{\ast}_{t})\|\gamma_E(\beta_t))\nonumber\\
    &=D(\rho_{SE}(t)\|\rho_{S}(t)\otimes \gamma_E(\beta^{\ast}_t))+D(\gamma_E(\beta^{\ast}_{t})\|\gamma_E(\beta_t)).
    \label{eq:Pythagorean_product}
\end{align}
Equation~\eqref{eq:Pythagorean_product} implies that the effective inverse temperature $\beta^{\ast}_t$ can be interpreted as the projection from the point $\rho_{SE}(t)$ onto the manifold $\{\rho_S(t)\otimes\gamma_E(\beta)\; |\;\beta\in\mathbb{R}\}$:
\begin{align}
    \beta^{\ast}_t=\mathrm{argmin}_{\beta\in\mathbb{R}} D(\rho_{SE}(t)\|\rho_{S}(t)\otimes \gamma_E(\beta)).
\end{align}
From a geometric perspective, $\Delta D[\beta^{\ast}_t]=0$ in Eq.~\eqref{eq:Delta_D_beta} follows because $\beta^{\ast}_t$ is a stationary point of $D(\rho_{SE}(t)\|\rho_{S}(t)\otimes \gamma_E(\beta))$.
By using Eq.~\eqref{eq:Pythagorean_product}, we obtain
\begin{align}
    &\Delta \Sigma(\beta_{0},\beta_{\tau})
    \nonumber\\
    &=\Delta \Sigma^{\ast}+D(\gamma_E(\beta^{\ast}_{\tau})\|\gamma_E(\beta_{\tau}))-D(\gamma_E(\beta^{\ast}_0)\|\gamma_E(\beta_0)).
    \label{eq:ent_general_definition}
\end{align}
This relation is a generalization of $\Delta\Sigma(\beta^{\ast}_{0},\beta^{\ast}_{0})=\Delta \Sigma^{\ast}+D(\gamma_E(\beta^{\ast}_{\tau})\|\gamma_E(\beta^{\ast}_0))$ which was shown in Ref.~\cite{strasberg2021first, strasberg2021clausius}. 
This relation implies that $\Delta \Sigma^\ast$ is the minimum among the entropy production for the effective temperature such that $\beta_0=\beta^{\ast}_0$:
\begin{align}
    \Delta\Sigma^\ast=\min_{\beta_\tau\in\mathbb{R}} \Delta\Sigma(\beta^\ast_0,\beta_\tau).
\end{align}
By using Eq.~\eqref{eq:D_decomp_mutual_DE}, $S(\rho_{SE})=S(\sigma_{SE})$ and the following relation (see Appendix~\ref{sec:Pythagorean}),
\begin{align}
    &D(\rho_{SE}(t)\|\rho_{S}(t)\otimes \gamma_E(\beta^{\ast}_{t}))\nonumber\\
    &=I(\rho_{SE}(t))+S(\gamma_E(\beta^{\ast}_{t}))-S(\rho_E(t))\nonumber\\
    &=S(\rho_S(t))+S(\gamma_E(\beta^{\ast}_t))-S(\rho_{SE}(t)),
    \label{eq:D_min_diff_ent}
\end{align}
$\Delta \Sigma^{\ast}$ can be written as follows~\cite{strasberg2021first}.
\begin{align}
    \Delta \Sigma^{\ast}=\Delta I+\Delta S_{\gamma}=\Delta S_S + \Delta S_E+\Delta S_{\gamma}.
\end{align}
Here we define $\Delta S_{\gamma}:=\{S(\gamma_E(\beta^{\ast}_{\tau}))-S(\sigma_{E})\}-\{S(\gamma_E(\beta^{\ast}_{0}))-S(\rho_{E})\}$.

We next consider the entropy production rate.
In the definition of Eq.~\eqref{eq:def_EP_geometric}, since no assumptions are imposed on the initial state, we can take $t=0$ at an arbitrary time. In the limit $\tau\rightarrow 0$ in Eq.~\eqref{eq:def_EP_geometric} and rewriting $\tau=0$ as $\tau=t$, from Eqs.~\eqref{eq:def_EP_time_variant_beta}, ~\eqref{eq:relation_clausius} and~\eqref{eq:Delta_D_beta}, we obtain
\begin{align}
    \sigma(t,\beta_t, \dot{\beta}_t)&:=\frac{d}{dt}D(\rho_{SE}(t)\|\rho_{S}(t)\otimes \gamma_E(\beta_{t}))\nonumber\\
    &=\frac{dS(\rho_S(t))}{dt}-\beta_t\frac{dQ(t)}{dt}\nonumber\\
    &+\dot{\beta_t} (\mathrm{tr}_E[\gamma_E(\beta^{\ast}_t)H_E]-\mathrm{tr}_E[\gamma_E(\beta_t)H_E]),
    \label{eq:EP_rate_expansion}
\end{align}
where $\sigma(t,\beta_t, \dot{\beta}_t)$ is the entropy production rate. 
As in conventional thermodynamics, if we require the third term on the right-hand side to be zero, $\dot{\beta_t} =0$ or $\beta_t=\beta^{\ast}_t$ must hold at all times. Here we use that $\mathrm{tr}_E[\gamma_E(\beta)H_E]$ is strictly decreasing with respect to $\beta$ (see Appendix~\ref{sec:uniqueness}).

\subsection{Lower bounds}

We consider the general effective temperature again.
For general initial states, the entropy production can be negative.
Suppose that the entropy production $\Delta \Sigma^{\ast}$ has a lower bound:
\begin{align}
    \Delta \Sigma^{\ast}&\geq  \Lambda(\rho_{SE}),
    \label{eq:illustrative_bound}
\end{align}
where $ \Lambda(\rho_{SE})$ is a lower bound comprising only an initial state. 
From Eq.~\eqref{eq:ent_general_definition}, a lower bound for $\Delta \Sigma(\beta_{0},\beta_{\tau})$ is given by
\begin{align}
    \Delta \Sigma(\beta_{0},\beta_{\tau})\geq \Lambda(\rho_{SE})-D(\gamma_E(\beta^{\ast}_{0})\|\gamma_E(\beta_0)).
    \label{eq:lowerbound_general_EP}
\end{align}
Since $D(\gamma_E(\beta^{\ast}_{0})\|\gamma_E(\beta_0))> 0$ for $\beta_0\neq \beta^{\ast}_0$,  the right-hand side gives the largest value  when $\beta_0=\beta^{\ast}_0$.
The definition~\eqref{eq:def_EP_geometric} for $\beta_t=\beta^{\ast}_t$ yields
\begin{align}
    \Delta \Sigma^{\ast}\geq -D(\rho_{SE}\|\rho_{S}\otimes \gamma_E(\beta^{\ast}_{0}))=:\Lambda_S(\rho_{SE}).
    \label{eq:EP_lb_relative_ent}
\end{align}
By applying Eq.~\eqref{eq:D_min_diff_ent} for $t=0$, we obtain the lower bound for the entropy production $\Delta \Sigma^{\ast}$:
\begin{align}
    \Lambda_S(\rho_{SE})&= S(\rho_{SE})-S(\rho_S)-S(\gamma_E(\beta^{\ast}_0))\nonumber\\
    &=S(\rho_{SE})-S(\rho_S\otimes\gamma_E(\beta^{\ast}_{0})).
    \label{eq:lb_main_result_ent}
\end{align}
Equations~\eqref{eq:lowerbound_general_EP} and~\eqref{eq:lb_main_result_ent} are the third main results of this paper. 
We comment the two special cases in order. When $\rho_E=\gamma_E(\beta)$, it follows that $\beta^{\ast}_0=\beta$ and 
\begin{align}
     \Lambda_S(\rho_{SE})=-I(\rho_{SE}),
\end{align}
where we use Eq.~\eqref{eq:def_mutual_info}. 
When $\rho_{SE}=\rho_{S}\otimes \rho_E$, substituting $S(\rho_{SE})=S(\rho_S)+S(\rho_E)$ into Eq.~\eqref{eq:lb_main_result_ent} yields 
\begin{align}
    \Lambda_S(\rho_{SE})= S(\rho_E)-S(\gamma_E(\beta^{\ast}_{0})).
\label{eq:lb_main_result_ent_prod}
\end{align}

We next show lower bounds via the trace distance. Let $\|\rho-\sigma\|_1:=\mathrm{tr}[\sqrt{(\rho-\sigma)^\dagger (\rho-\sigma)}]$, and let $\mathcal{T}(\rho,\sigma):=\|\rho-\sigma\|_1/2$ be the trace distance. 
Letting $H_2(p):=-p\ln p-(1-p)\ln(1-p)$ be the two-dimensional Shannon entropy, the Fannes–Audenaert inequality~\cite{fannes1973continuity,audenaert2007sharp} is given by 
\begin{align}
    |S(\rho)-S(\sigma)|\le \mathcal{T}(\rho,\sigma) \ln (d-1)+H_2(\mathcal{T}(\rho,\sigma)),
    \label{eq:Fannes_Audenaert}
\end{align}
where $d$ is the dimension of $\rho$ and $\sigma$. This bound is known to be tight.
From Eqs.~\eqref{eq:EP_lb_relative_ent} and~\eqref{eq:lb_main_result_ent}, we obtain
\begin{align}
    &\Lambda_S(\rho_{SE}) \geq\Lambda_{\mathcal{T}}(\rho_{SE}):=-\delta_{\mathcal{T}}\ln (d_Sd_E-1)-H_2(\delta_{\mathcal{T}}) ,
    \label{eq:lb_trace}
\end{align}
where $\delta_{\mathcal{T}}:=\mathcal{T}(\rho_{SE}, \rho_S\otimes\gamma_E(\beta^{\ast}_{0}))$.

Suppose that $\rho_{SE}$ can be written as 
\begin{align}
    \rho_{SE}=\rho_S\otimes\gamma_E(\beta)+\chi_{SE}.
    \label{eq:init_chi_SE}
\end{align}
Here $\chi_{SE}$ is a Hermitian operator that satisfies
\begin{align}
    \mathrm{tr}_E[\chi_{SE}]&=0,\nonumber\\
    \mathrm{tr}_S[\chi_{SE}]&=\chi_E, 
\end{align}
and $\chi_E$ is a Hermitian operator that has no diagonal elements when represented in the energy eigenbasis of $H_E$. From the assumption for $\chi_E$,  Eq.~\eqref{eq:inverse_temp_condition} yields $\beta^{\ast}_0=\beta$ and it follows that $\delta_{\mathcal{T}}=\|\chi_{SE}\|_1/2$. 
Thus, we can calculate the lower bound Eq.~\eqref{eq:lb_trace} from $\|\chi_{SE}\|_1$. 
When $\rho_{SE}=\rho_S\otimes \rho_E$, from  $\|A\otimes B\|_1=\|A\|_1\|B\|_1$  
and $\|\rho_S\|_1=1$, 
it follows that $\delta_{\mathcal{T}}=\mathcal{T}(\rho_{E}, \gamma_E(\beta^{\ast}_{0}))$. By combining 
Eq.~\eqref{eq:lb_main_result_ent_prod} with Eq.~\eqref{eq:Fannes_Audenaert}, we obtain
\begin{align}
    &\Lambda_S(\rho_{SE}) \geq\Lambda^\prime_{\mathcal{T}}(\rho_{SE}):=-\delta_{\mathcal{T}}\ln (d_E-1)-H_2(\delta_{\mathcal{T}}) , \nonumber\\
    &\text{if $\rho_{SE}=\rho_S\otimes\rho_E$.} 
    \label{eq:lb_trace_prod}
\end{align}

\subsection{Sufficient condition for nonnegative value}

We show sufficient conditions for $\Delta \Sigma(\beta_0, \beta_{\tau})$ to be nonnegative via the trace distance. This relation is useful for visually understanding the region where entropy production is nonnegative in Section~\ref{eq:subsec_two_dimension}.
By combining the quantum Pinsker inequality~\cite{OhyaPetz:2004:QuantumEntropy} $D(\rho\|\sigma)\geq 2\mathcal{T}(\rho,\sigma)^2$ with Eqs.~\eqref{eq:def_EP_geometric}, ~\eqref{eq:lowerbound_general_EP} and~\eqref{eq:lb_trace}, it follows that
\begin{align}
    &\Delta \Sigma(\beta_{0},\beta_{\tau})\geq 2\mathcal{T}(\sigma_{SE},\sigma_S\otimes\gamma_E(\beta_{\tau}))^2+\Lambda_{\mathcal{T}}(\rho_{SE})-D_\gamma(\beta_0),
    \label{eq:inequality_trace_dist}
\end{align}
where $D_\gamma(\beta_{0}):=D(\gamma_E(\beta^{\ast}_{0})\|\gamma_E(\beta_0))$.
Thus, we obtain the sufficient condition for the entropy production to be nonnegative as follows.
\begin{align}
    &\mathcal{T}(\sigma_{SE},\sigma_S\otimes\gamma_E(\beta_{\tau}))^2\geq \frac{1}{2}\left(D_\gamma(\beta_{0})-\Lambda_{\mathcal{T}}(\rho_{SE})\right).
    \label{eq:sufficient_cond}
\end{align}

In the following, let us consider the initial state to be in product form, $\rho_{SE}=\rho_S\otimes\rho_E$. 
From Eq.~\eqref{eq:D_decomp_mutual_DE}, we obtain 
$D(\sigma_{SE}\|\sigma_{S}\otimes \gamma_E(\beta_{\tau}))\geq D(\sigma_{E}\|\gamma_E(\beta_{\tau}))$.
As in Eq.~\eqref{eq:inequality_trace_dist}, from Eq.~\eqref{eq:lb_trace_prod}, we obtain
\begin{align}
    \Delta \Sigma(\beta_{0},\beta_{\tau})&\geq D(\sigma_{E}\|\gamma_E(\beta_{\tau}))+\Lambda^\prime_{\mathcal{T}}(\rho_{SE})- D_\gamma(\beta_{0})\nonumber\\
    &\geq 2\mathcal{T}(\sigma_{E},\gamma_E(\beta_{\tau}))^2+\Lambda^\prime_{\mathcal{T}}(\rho_{SE})- D_\gamma(\beta_{0}).
\end{align}
Thus, we obtain the following sufficient condition:
\begin{align}
    \mathcal{T}(\sigma_{E},\gamma_E(\beta_{\tau}))^2\geq \frac{1}{2}\left(D_\gamma(\beta_{0})-\Lambda^\prime_{\mathcal{T}}(\rho_{SE})\right).
    \label{eq:sufficient_cond_prod}
\end{align}

\subsubsection{Example \label{eq:subsec_two_dimension}}

We consider a two-dimensional environment. Let $[0,\epsilon]$ be the eigenvalues of the Hamiltonian of the environment $H_E$. Let $\mathcal{M}_E$ be a set of all possible density operators of the environment.
We choose the eigenvectors of energy as the basis of the Hilbert space of the environment.
For $p,s\in\mathbb{R}$ and $\mathfrak{a},\mathfrak{b}\in \mathbb{C}$, the density operators of the environment are given by
\begin{align}
    \rho_E=\frac{1}{2}
    \begin{bmatrix}
        1+p& \mathfrak{a} \\
        \mathfrak{a}^* & 1-p \\
    \end{bmatrix}
    ,
\end{align}
and
\begin{align}
    \sigma_E=\frac{1}{2}
    \begin{bmatrix}
        1+s& \mathfrak{b} \\
        \mathfrak{b}^* & 1-s \\
    \end{bmatrix}
    ,
\end{align}
where $\mathfrak{a}^\ast$ and $\mathfrak{b}^\ast$ denote the complex conjugate. 
Since the characteristic equation for $\sigma_E$ is $\lambda^2-\lambda+(1-s^2-|\mathfrak{b}|^2)/ 4 = 0$, for eigenvalues to exist in $[0,1]$, $s^2+|\mathfrak{b}|^2 \le 1$ must hold. Thus, we identify the set $\mathcal{M}_E$ with the three-dimensional ball. 
Since the Gibbs states are diagonalized by the eigenstates of energy, they are distributed over $\mathfrak{b}=0$ ($s$-axis). For $r(\beta)\in\mathbb{R}$, the Gibbs state is given by 
\begin{align}
    \gamma_E(\beta)=\frac{1}{2}
    \begin{bmatrix}
        1+r(\beta)& 0 \\
        0 & 1-r(\beta) \\
    \end{bmatrix}.
    \label{eq:matrix_gamma_E}
\end{align}
Here, the negative $r(\beta)$ corresponds to the negative effective temperature.
From Eq.~\eqref{eq:inverse_temp_condition}, we obtain $r(\beta^{\ast}_0)=p$ and $r(\beta^{\ast}_{\tau})=s$. Since the eigenvalues of $\rho_E-\gamma_E(\beta^{\ast}_0)$ and $\sigma_E-\gamma_E(\beta_{\tau})$ are $\pm |\mathfrak{a}|/2$ and $\pm \sqrt{(s-r(\beta_{\tau}))^2+|\mathfrak{b}|^2}/2$, respectively, the trace distances are given by $\delta_{\mathcal{T}}=|\mathfrak{a}|/2$ and $\mathcal{T}(\sigma_E,\gamma_E(\beta_{\tau}))=\sqrt{(s-r(\beta_{\tau}))^2+|\mathfrak{b}|^2}/2$.
Substituting these results into Eq.~\eqref{eq:sufficient_cond_prod} and using Eq.~\eqref{eq:lb_trace_prod}, a sufficient condition for the entropy production to be nonnegative is given by
\begin{align}
    (s-r(\beta_{\tau}))^2+|\mathfrak{b}|^2\geq 2H_2\left(\frac{|\mathfrak{a}|}{2}\right)+2 D_\gamma(\beta_{0}).
    \label{eq:sufficient_cond_two_dim}
\end{align}
The right-hand side depends only on the non-diagonal element of $\rho_E$ and the difference between the Gibbs states at $\beta_0$ and $\beta^{\ast}_0$. When  $\beta_{t} =\beta^C=\beta_{0}$,  this is the region outside the ball with center $(r(\beta_{0}),0,0)$ and radius $\sqrt{2H_2\left(|\mathfrak{a}|/2\right)+2D_\gamma(\beta_{0})}$. When $\beta_{t} = \beta^{\ast}_{t}$, recall that $r(\beta^{\ast}_{\tau})=s$, Eq.~\eqref{eq:sufficient_cond_two_dim} reduces to 
\begin{align}
    |\mathfrak{b}|^2\geq 2H_2\left(\frac{|\mathfrak{a}|}{2}\right).
    \label{eq:sufficient_cond_two_dim_tilde}
\end{align}

\section{Conclusion}
In this study, we have proposed the definition of entropy production with general effective temperatures via the difference of the quantum relative entropy. This definition is a natural extension of the entropy production for a large environment, imposing the requirement that 1) the second law of thermodynamics holds for the initial state in product form and the initial state of the environment is taken as the Gibbs state, 2) entropy production becomes zero when the initial and the final state are identical including the effective temperature. The proposed entropy production can be decomposed into a Clausius-type component and a component varying with time dependence of the effective temperature. We have shown that our definition reduces to the previously proposed Clausius-type definition for the effective temperature $\beta^{\ast}_t$, which is defined such that the instantaneous mean energy of the environment agrees with that of a Gibbs state. The entropy production with general effective temperature is decomposed as 
For the entropy production rate to be the sum of the change in entropy and the change in heat as in conventional thermodynamics, the effective temperature must be either constant or equal to $\beta^{\ast}_t$.
The entropy production for general effective temperatures can be decomposed into the sum of the entropy production for $\beta^{\ast}_t$ and the quantum relative entropy arising from the difference in effective temperature definitions.
By using this result, We have also derived lower bounds on the entropy production for general initial states, and we have shown a sufficient condition for non-negative entropy production based on the trace distance. 
Overall, this study has contributed to the understanding of the entropy production with the effective temperature.

\begin{acknowledgments}

This work was supported by JSPS KAKENHI Grant Numbers JP23K24915 and JP24K03008.

\end{acknowledgments}

\appendix
\begin{widetext}

\section{Uniqueness of solution of Eq.~\eqref{eq:inverse_temp_condition} \label{sec:uniqueness}}
Differentiating $\mathbb{E}_\gamma[H_E](\beta):=\mathrm{tr}_E[\gamma_E(\beta) H_E]$ with respect to $\beta$ yields
\begin{align}
    \frac{d}{d\beta}\mathbb{E}_\gamma[H_E](\beta)=-\mathrm{Var}_\gamma[H_E](\beta)<0,
    \label{eq:monotonicity}
\end{align}
where $\mathrm{Var}_\gamma[H_E](\beta):=\mathbb{E}_\gamma[H_E^2](\beta)-\mathbb{E}_\gamma[H_E](\beta)^2$ is the variance of $H_E$. Let $\epsilon_{\max}$ and $\epsilon_{\min}$ be maximum and minimum eigenvalue of $H_E$, respectively. Since $\mathbb{E}_\gamma[H_E](\beta)$ is strictly decreasing and $\mathbb{E}_\gamma[H_E](-\infty)=\epsilon_{\max}$, $\mathbb{E}_\gamma[H_E](\infty)=\epsilon_{\min}$, equation~\eqref{eq:inverse_temp_condition} always has a unique solution.

\section{Proof of Eq.~\eqref{eq:relation_clausius} \label{sec:relation_clausius}}
Substituting $S(\sigma_{SE})=S(\rho_{SE})$ into Eq.~\eqref{eq:def_EP_geometric}, we obtain
\begin{align}
    \Delta \Sigma(\beta_{0},\beta_{\tau})&= \Delta S_S + \Delta S_E + D(\sigma_E\|\gamma_E(\beta_{\tau}))-D(\rho_E\|\gamma_E(\beta_0))\nonumber\\
    &=\Delta S_S+\beta_{\tau}\mathrm{tr}_E[\gamma_E(\beta_{\tau}) H_E]-\beta_0\mathrm{tr}_E[\gamma_E(\beta_{0}) H_E]+\ln \mathcal{Z}(\beta_{\tau})-\ln \mathcal{Z}(\beta_0)\nonumber\\
    &=\Delta S_S+\int_0^\tau \frac{d}{dt}\left(\beta_t \mathrm{tr}_E[\rho_{E}(t)H_E]+\ln \mathcal{Z}(\beta_t)\right)dt\nonumber\\
    &=\Delta \Sigma_{\mathrm{CL}}[\beta_{t}]+\int_0^\tau \left(\frac{d\beta_t }{dt}\mathrm{tr}_E[\rho_{E}(t)H_E]+\frac{d}{dt}\ln \mathcal{Z}(\beta_t)\right)\nonumber\\
    &=\Delta \Sigma_{\mathrm{CL}}[\beta_{t}]+\int_0^\tau \dot{\beta_t} \left.\frac{\partial}{\partial\beta} D(\rho_E(t)\| \gamma_E(\beta)) \right|_{\beta=\beta_{t}},
    \label{eq:DeltaD_Delta_Sigma}
\end{align}
where we use 
\begin{align}
    D(\rho_E(t)\| \gamma_E(\beta))=-S(\rho_E(t))+\beta \mathrm{tr}_E[\rho_{E}(t)H_E]+\ln \mathcal{Z}(\beta),
    \label{eq:relative_EP_expansion}
\end{align}
in the last equality.
By using Eq.~\eqref{eq:D_decomp_mutual_DE}, we obtain Eq.~\eqref{eq:relation_clausius} for 
\begin{align}
    \Delta D[\beta_t]&=\int_0^\tau \dot{\beta_t} \left.\frac{\partial}{\partial\beta} D(\rho_E(t)\| \gamma_E(\beta)) \right|_{\beta=\beta_{t}}=\int_0^\tau \dot{\beta_t} \left.\frac{\partial}{\partial\beta} D(\rho_{SE}(t)\|\rho_{S}(t)\otimes \gamma_E(\beta))\right|_{\beta=\beta_{t}}dt.
\end{align}
By using Eqs.~\eqref{eq:inverse_temp_condition}, ~\eqref{eq:relative_EP_expansion} and $\partial_\beta \ln \mathcal{Z}(\beta)=-\mathrm{tr}_E[\gamma_E(\beta)H_E]$, we obtain
\begin{align}
    \left.\frac{\partial}{\partial\beta} D(\rho_E(t)\| \gamma_E(\beta)) \right|_{\beta=\beta_{t}}&=\mathrm{tr}_E[\rho_E(t)H_E]-\mathrm{tr}_E[\gamma_E(\beta_t)H_E]\nonumber\\
    &=\mathrm{tr}_E[\gamma_E(\beta^{\ast}_t)H_E]-\mathrm{tr}_E[\gamma_E(\beta_t)H_E].
\end{align}
Therefore, it follows that the second equality in Eq.~\eqref{eq:Delta_D_beta}.

\section{Proof of Eq.~\eqref{eq:Pythagorean_eq} and~\eqref{eq:D_min_diff_ent}  \label{sec:Pythagorean}}
The Gibbs state $\gamma_E(\beta^{\ast}_{t})$ satisfies
\begin{align}
    &0=D(\gamma_E(\beta^{\ast}_{t}) \| \gamma_E(\beta^{\ast}_{t}))=D\left(\gamma_E(\beta^{\ast}_{t}) \| e^{-{\beta^{\ast}_t H_E}}/\mathcal{Z}(\beta^{\ast}_{t})\right)\nonumber\\
    &=-S(\gamma_E(\beta^{\ast}_{t}))+\beta^{\ast}_{t}\mathrm{tr}_E[H_E \gamma_E(\beta^{\ast}_{t})]+\ln \mathcal{Z}(\beta^{\ast}_{t}).
    \label{eq:ent_energy_relation}
\end{align}
By combining Eq.~\eqref{eq:relative_EP_expansion} with the definition of $\beta^{\ast}_t$ [Eq.~\eqref{eq:inverse_temp_condition}] and using this equality, we obtain
\begin{align}
    &D(\rho_E(t) \| \gamma_E(\beta^{\ast}_{t}))=-S(\rho_E(t))+\beta^{\ast}_{t}\mathrm{tr}_E[H_E \gamma_E(\beta^{\ast}_{t})]+\ln \mathcal{Z}(\beta^{\ast}_{t})\nonumber\\
    &=S(\gamma_E(\beta^{\ast}_{t}))-S(\rho_E(t)).
    \label{eq:D_E_ent_diff}
\end{align}
From Eqs.~\eqref{eq:inverse_temp_condition},~\eqref{eq:relative_EP_expansion} and this relation, we obtain Eq.~\eqref{eq:Pythagorean_eq} as follows.
\begin{align}
    &D(\rho_E(t) \| \gamma_E(\beta))=-S(\rho_E(t))+\beta\mathrm{tr}_E[H_E \gamma_E(\beta^{\ast}_{t})]+\ln \mathcal{Z}(\beta)\nonumber\\
    &=[S(\gamma_E(\beta^{\ast}_{t}))-S(\rho_E(t))]+[-S(\gamma_E(\beta^{\ast}_{t}))+\beta\mathrm{tr}_E[H_E \gamma_E(\beta^{\ast}_{t})]+\ln \mathcal{Z}(\beta)]\nonumber\\
    &=D(\rho_E(t) \| \gamma_E(\beta^{\ast}_{t}))+D(\gamma_E(\beta^{\ast}_{t})\|\gamma_E(\beta)).
    \label{eq:Pythagorean_eq_base}
\end{align}
Substituting Eq.~\eqref{eq:D_E_ent_diff} and the definition of the mutual information [Eq.~\eqref{eq:def_mutual_info}] into Eq.~\eqref{eq:D_decomp_mutual_DE}, we obtain Eq.~\eqref{eq:D_min_diff_ent}:
\begin{align}
    &D(\rho_{SE}(t)\|\rho_{S}(t)\otimes \gamma_E(\beta^{\ast}_{t}))=I(\rho_{SE}(t))+D(\rho_E(t)\|\gamma_E(\beta^{\ast}_t))\nonumber\\
    &=I(\rho_{SE}(t))+S(\gamma_E(\beta^{\ast}_{t}))-S(\rho_E(t))=S(\rho_S(t))+S(\gamma_E(\beta^{\ast}_{t}))-S(\rho_{SE}(t)).
    \label{eq:D_SE_ent_Dgamma}
\end{align}

\end{widetext}

\end{document}